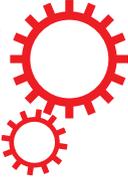

# SCIENTIFIC REPORTS

**OPEN** Magnetic coherent tunnel junctions with periodic grating barrier

Henan Fang[1], Mingwen Xiao[2], Wenbin Rui[2], Jun Du[2,3] & Zhikuo Tao[1]

A new spintronic theory has been developed for the magnetic tunnel junction (MTJ) with single-crystal barrier. The barrier will be treated as a diffraction grating with intralayer periodicity, the diffracted waves of tunneling electrons thus contain strong coherence, both in charge and especially in spin. The theory can answer the two basic problems present in MgO-based MTJs: (1) Why does the tunneling magnetoresistance (TMR) oscillate with the barrier thickness? (2) Why is the TMR still far away from infinity when the two electrodes are both half-metallic? Other principal features of TMR can also be explained and reproduced by the present work. It also provides possible ways to modulate the oscillation of TMR, and to enhance TMR so that it can tend to infinity. Within the theory, the barrier, as a periodic diffraction grating, can get rid of the confinement in width, it can vary from nanoscale to microscale. Based on those results, a future-generation MTJ is proposed where the three pieces can be fabricated separately and then assembled together, it is especially appropriate for the layered materials, e.g., $MoS_2$ and graphite, and most feasible for industries.

The tunneling magnetoresistance (TMR) was first studied theoretically and observed experimentally by Jullière in 1975 with magnetic tunneling junctions (MTJs) at low temperatures[1]. However, it can hardly be observed at room temperature until the MTJs with amorphous aluminum oxide (Al-O) barrier were fabricated in 1995[2,3]. From then on, the TMR effect has been applied in magnetic sensors and memory devices, and thus received considerable attention for the last twenty years[4–6]. Nevertheless, the Al-O-based MTJs can only exhibit a TMR ratio up to 80% because the Al-O barrier is amorphous and thus give rise to strong incoherent tunneling process[6]. This low TMR ratio seriously limits the feasibility of spintronics devices[7]. Butler et al.[8] predicted theoretically that, if MgO is used to prepare the MTJ barrier, the TMR can acquire a very high value. The prediction was verified soon by S. S. P. Parkin et al.[9] and S. Yuasa et al.[10]. Since then, the MgO-based MTJs have been widely investigated over the last decade[11–19].

Apart from the high TMR ratio, the MgO-based MTJs manifest many novel physical properties because the MgO barrier can be prepared into an ultrathin single-crystal film, especially by using molecular beam epitaxy (MBE)[10,16]. Of those novel properties, it is the most distinguished and puzzling that the parallel resistance ($R_P$), antiparallel resistance ($R_{AP}$), and TMR all oscillate with the barrier thickness, which is radically different from the case of the conventional MTJs with amorphous Al-O barrier where no oscillation is found[20,21]. The oscillation was first observed by Yuasa et al.[10] in the study of a series of Fe(001)/MgO(001)/Fe(001) MTJs with barrier thickness varying from 1.2 nm to 3.2 nm, and then reproduced by many research groups[16–19]. The experiments[10,16–19] reveal that $R_P$ always exhibits only one period of oscillation. For $R_{AP}$ and TMR, the situation is somewhat different: They always exhibit a basic period of oscillation, too; but in some cases they can exhibit a secondary period of oscillation which is longer than the basic one, as reported in refs 16,19. Furthermore, the experiments indicate that the basic periods for $R_P$, $R_{AP}$, and TMR are nearly the same. Finally, the amplitude of TMR is discovered to be 20% ~ 40% when the long period occurs, it becomes about 100% otherwise. Quite regrettably, the TMR oscillations can neither be explained by the above theory proposed by Butler et al., just as pointed out in refs 16,18, nor by other ordinary methods, such as ab-initio band structure[22], Landauer formula[23], and non-equilibrium Green's function quantum transport calculations[24,25]. The physical mechanism for those oscillations has not been clarified as yet, to our knowledge.

In order to interpret this puzzle, we would like to present a microscopic theory for the MTJ with single-crystal barrier. In this theory, the tunneling process will be regarded as the scattering of the electron wave by the periodic potential of the barrier. Physically, such scattering is identical to the diffraction of light through an optical grating.

[1]Nanjing University of Posts and Telecommunications, Nanjing 210023, China. [2]Department of Physics, Nanjing University, Nanjing 210093, China. [3]Collaborative Innovation Center of Advanced Microstructures, Nanjing University, Nanjing 210093, China. Correspondence and requests for materials should be addressed to M.X. (email: xmw@nju.edu.cn)

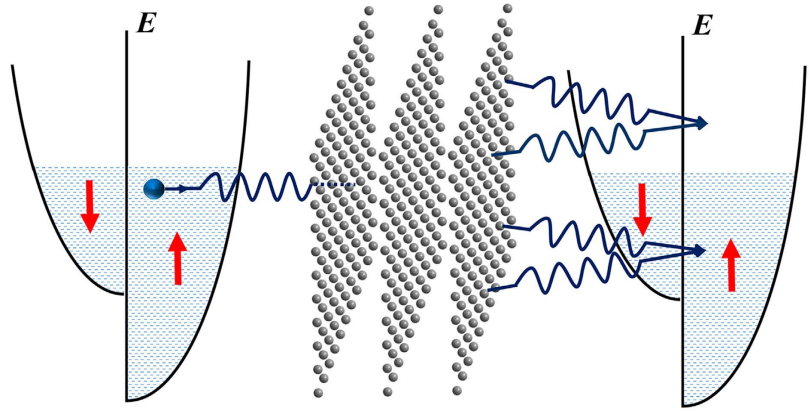

**Figure 1. The diagrammatic sketch of the diffraction process where the incident electron is spin-up.**

After the diffraction, the current of electrons, as the total scattering section, has the effect of space coherence. As will be seen in the following, it is this effect of space coherence that leads to the oscillations of $R_P$, $R_{AP}$ and TMR with the barrier thickness.

To begin with, let us consider a MTJ consisting of a thin single-crystal barrier. Physically, we will treat the barrier as a periodic potential, instead of the trapezoidal one as is used in the previous works[26–30] where Al-O-based MTJs are involved. For the Al-O-barrier, it is amorphous so that the barrier potential is statistically smoothed by the disorders in the barrier. It is, therefore, rational to model the Al-O-barrier as a trapezoidal potential. But now, the barrier is single-crystal, the periodicity becomes fundamentally important because a periodic structure will cause strong effect of coherence to the electrons passing through it. As to the ferromagnetic (FM) electrodes, we will treat them, as usual, with the free-electron model[26–30].

Suppose that the atomic potential of the barrier is $v(\mathbf{r})$, and that the total number of the layers of the barrier is $n$. Then, the periodic potential $U(\mathbf{r})$ of the barrier can be written as

$$U(\mathbf{r}) = \sum_{l_3=0}^{n-1} \sum_{\mathbf{R}_h} v(\mathbf{r} - \mathbf{R}_h - l_3 \mathbf{a}_3), \qquad (1)$$

where $\mathbf{R}_h$ is a two-dimensional lattice vector of the barrier: $\mathbf{R}_h = l_1 \mathbf{a}_1 + l_2 \mathbf{a}_2$, with $\mathbf{a}_1$ and $\mathbf{a}_2$ being the primitive vectors of the atomic layers, and $l_1$ and $l_2$ the corresponding integers. The $\mathbf{a}_3$ is the third primitive vector of the barrier, with $l_3$ the corresponding integer. Letting $\mathbf{e}_z = \mathbf{a}_1 \times \mathbf{a}_2 / |\mathbf{a}_1 \times \mathbf{a}_2|$, we shall set $\mathbf{e}_z$ point from the upper electrode to the lower one, which is antiparallel to the direction of the tunneling current.

Now, let us consider the case that the two FM electrodes are magnetically parallel. Suppose that a spin-up electron tunnels from the upper FM electrode into the lower one and occupies the spin-up state. The incident electron can be described by the plane wave as follows,

$$\psi_{i\uparrow} = \exp(i\mathbf{k} \cdot \mathbf{r}), \qquad (2)$$

where $\mathbf{k}$ denotes the wave vector. Physically, this incident wave will be diffracted coherently by the periodic potential of the barrier. As a result, the out-going waves arriving at the lower FM electrode, i.e., the so-called transmitted waves, will manifest strong effects of coherence. In order to elucidate the effects of coherence, we would employ the Bethe theory and the two-beam approximation, which are standard methods for the Laue case of transmission through a perfect, parallel-sided crystal plate[31]. According to those methods, the transmitted wave function $\psi_\uparrow(\mathbf{r})$ can be obtained as follows,

$$\psi_\uparrow(\mathbf{r}) = \frac{1}{2}(e^{i\mathbf{p}_+\cdot\mathbf{r}} + e^{i\mathbf{p}_-\cdot\mathbf{r}} + e^{i\mathbf{q}_+\cdot\mathbf{r}} - e^{i\mathbf{q}_-\cdot\mathbf{r}}), \qquad (3)$$

where

$$\mathbf{p}_\pm = \mathbf{k}_h + [\mathbf{k}^2 - \mathbf{k}_h^2 \pm 2m\hbar^{-2}v(\mathbf{K}_h)]^{1/2} \mathbf{e}_z, \qquad (4)$$

$$\mathbf{q}_\pm = \mathbf{k}_h + \mathbf{K}_h + [\mathbf{k}^2 - (\mathbf{k}_h + \mathbf{K}_h)^2 \pm 2m\hbar^{-2}v(\mathbf{K}_h)]^{1/2} \mathbf{e}_z. \qquad (5)$$

Schematically, this diffraction process can be illustrated as in Fig. 1. In deriving equation (3), we have supposed that the two FM electrodes are the same. Here, $\mathbf{k}_h$ is the normal projection of the incident wave vector $\mathbf{k}$ on the plane spanned by $\mathbf{a}_1$ and $\mathbf{a}_2$. As to $v(\mathbf{K}_h)$, it represents the Fourier transform of $v(\mathbf{r})$: $v(\mathbf{K}_h) = \Omega^{-1} \int d\mathbf{r} v(\mathbf{r}) e^{-i\mathbf{K}_h\cdot\mathbf{r}}$ where $\Omega$ is the volume of the primitive cell of the barrier: $\Omega = (\mathbf{a}_1 \times \mathbf{a}_2)\cdot\mathbf{a}_3$, and $\mathbf{K}_h$ a planar vector reciprocal to the intralayer lattice vectors $\mathbf{R}_h$. Physically, each incident wave vector $\mathbf{k}$ will correspond to one reciprocal vector $\mathbf{K}_h$ such that the magnitude of the planar vector $\mathbf{k}_h + \mathbf{K}_h$ is minimal. By using the out-going wave $\psi_\uparrow(\mathbf{r})$, the transmission coefficient for the spin-up to spin-up tunneling can be calculated as follows,

$$\begin{aligned}
T_{\uparrow\uparrow}(\mathbf{k}) &= \frac{|\mathbf{a}_1 \times \mathbf{a}_2|}{2ik_z S_h} \iint dx_1 dx_2 \left[ \psi_\uparrow^*(\mathbf{r}) \frac{\partial}{\partial z} \psi_\uparrow(\mathbf{r}) - \text{c.c.} \right] \\
&= \frac{1}{8k_z} \Big\{ p_+^z e^{i[p_+^z - (p_+^z)^*]d} + p_-^z e^{i[p_-^z - (p_-^z)^*]d} + q_+^z e^{i[q_+^z - (q_+^z)^*]d} + q_-^z e^{i[q_-^z - (q_-^z)^*]d} \\
&\quad + \left[ p_+^z e^{i[p_+^z - (p_-^z)^*]d} + p_-^z e^{i[p_-^z - (p_+^z)^*]d} - q_+^z e^{i[q_+^z - (q_-^z)^*]d} - q_-^z e^{i[q_-^z - (q_+^z)^*]d} \right] + \text{c.c.} \Big\}
\end{aligned} \quad (6)$$

where $k_z = \mathbf{k} \cdot \mathbf{e}_z$, $p_\pm^z = \mathbf{p}_\pm \cdot \mathbf{e}_z$, $q_\pm^z = \mathbf{q}_\pm \cdot \mathbf{e}_z$, $\mathbf{r} = x_1 \mathbf{a}_1 + x_2 \mathbf{a}_2 + d\mathbf{e}_z$ with $d$ being the width of the barrier, and $S_h$ is the cross sectional area of the barrier. To perform the integral above, we have employed the so-called Born-von Karman boundary condition. It is easy to know that the term $\exp(i[p_+^z - (p_-^z)^*]d)$ arises physically from the interference between the component waves $\exp(i\mathbf{p}_+ \cdot \mathbf{r})$ and $\exp(i\mathbf{p}_- \cdot \mathbf{r})$, it will oscillate with the barrier width $d$ if $p_+^z$ and $p_-^z$ are both real. Similar statements hold for the other three terms in the square brackets of equation (6). It will be seen in the following that it is just those interference terms that are responsible for the oscillations of $R_P$, $R_{AP}$, and TMR. From $T_{\uparrow\uparrow}$, the conductance $G_{\uparrow\uparrow}$ of zero bias voltage at zero temperature can be written as[26–28]

$$G_{\uparrow\uparrow} = \frac{e^2}{16\pi^3 \hbar} \int_0^{\pi/2} d\theta \int_0^{2\pi} d\varphi k_{F\uparrow}^2 \sin(2\theta) T_{\uparrow\uparrow}(k_{F\uparrow}, \theta, \varphi), \quad (7)$$

where $e$ denotes the electron charge, $\theta$ the angle between $\mathbf{k}$ and $\mathbf{e}_z$, $\varphi$ the angle between $\mathbf{k}_h$ and $\mathbf{a}_1$, and $k_{F\uparrow}$ the Fermi wave vector of the spin-up electrons:

$$k_{F\uparrow} = \sqrt{2m\hbar^{-2}(\mu + \Delta)} \quad (8)$$

with $m$ being the electron mass, and $\mu$ and $\Delta$ the chemical potential and half the exchange splitting of the FM electrodes, respectively. The other three conductances, $G_{\uparrow\downarrow}$, $G_{\downarrow\uparrow}$ and $G_{\downarrow\downarrow}$, can be obtained similarly. With them, one can obtain $G_P = G_{\uparrow\uparrow} + G_{\downarrow\downarrow}$, $G_{AP} = G_{\uparrow\downarrow} + G_{\downarrow\uparrow}$, $R_P = G_P^{-1}$, $R_{AP} = G_{AP}^{-1}$ and TMR $= G_P/G_{AP} - 1 = R_{AP}/R_P - 1$.

From now on, we shall apply the above formalism to the case of MgO-based MTJs. The lattice of MgO crystal is simple cubic, viz., $\mathbf{a}_1 \perp \mathbf{a}_2$, $\mathbf{a}_2 \perp \mathbf{a}_3$, $\mathbf{a}_3 \perp \mathbf{a}_1$ and $a_1 = a_2 = a_3$, therefore, $\mathbf{e}_z \parallel \mathbf{a}_3$. It can be easily seen that $(Q_h, 0, 0)$ should be chosen for the incident vector $\mathbf{k}$ with $\varphi \in [-\pi/4, \pi/4]$ where $Q_h = 2\pi/a_1$. Analogously, the $(0, Q_h, 0)$, $(-Q_h, 0, 0)$ and $(0, -Q_h, 0)$ should be chosen for the $\mathbf{k}$ with $\varphi \in [\pi/4, 3\pi/4]$, $[3\pi/4, 5\pi/4]$ and $[5\pi/4, 7\pi/4]$, respectively. Due to the planar symmetry, the contributions to the tunneling current from the four intervals of the angle $\varphi$ are the same. Therefore, it is enough for us to consider the interval $[-\pi/4, \pi/4]$. As a result, there are only four model parameters needed for the application of the present theory to the MgO-based MTJs: the magnitude of the reciprocal vector $\mathbf{K}_h$ ($\mathbf{K}_h = (Q_h, 0, 0)$), the Fourier transform of the periodic potential of the barrier $v(\mathbf{K}_h)$, the chemical potential $\mu$, and half the exchange splitting $\Delta$ of the FM electrodes. According to the data in ref. 32, $K_h = 2\pi/a_1 = 2.116 \times 10^{10}$ m$^{-1}$. In order to determine $v(\mathbf{K}_h)$, let us first discuss which property of material the parameter $v(\mathbf{K}_h)$ can be correlated with. As pointed out in ref. 31, the two-beam model is almost exactly the same as the nearly free electron approximation for solids, the main difference between them lies in that the aim of the former is to establish the wave vectors and amplitudes of the diffracted beams rather than to establish the energy levels of the system as does in the latter. What is the most important is that both of them contain $v(\mathbf{K}_h)$. As well-known, $v(\mathbf{K}_h)$ is proportional to the energy gap of bands in the nearly free electron approximation. That is to say, the parameter $v(\mathbf{K}_h)$ should be approximately proportional to the energy gap, in physics. Since the energy gap of MgO crystal is about 8 eV[33], we shall set $v(\mathbf{K}_h) = 16$ eV in this paper for the comparison of the theory with the experiments. As to the other two parameters, $\mu$ and $\Delta$, they are both independent of the MgO barrier, and thus should be determined by the FM electrodes. For different FM electrodes, they will get different values. The typical case is Fe electrodes, its $\mu$ is 11.1 eV[34]. As to the $\Delta$, it can be figured out from the $\mu$ and the spin polarization by using Hubbard model. According to the data of refs 35–37, it can be estimated to be about 8.2 eV, i.e., $\Delta \approx 8.2$ eV. Therefore, we shall in this paper let the chemical potential $\mu$ take on values from 10 eV to 13 eV, and the exchange splitting $\Delta$ take on values from 7 eV to 10 eV, so as to compare the theoretical results with the experiments on MTJs with different FM electrodes.

First, we would like to investigate the dependence of $R_P$, $R_{AP}$ and TMR on the thickness $d$ of MgO barrier. The results are shown in Figs 2 and 3 where $\mu$ and $\Delta$ are fixed respectively to be 11 eV and 9 eV. In Fig. 2, $\Delta$ varies from 7 eV to 10 eV. By contrast, $\mu$ will vary from 10 eV to 13 eV in Fig. 3. Here, the barrier width $d$ has been extended from discrete values to continuous real numbers through equation (6). As stated above, the most fundamental feature discovered by the experiments is that all the $R_P$, $R_{AP}$ and TMR oscillate with the barrier thickness. Both Figs 2 and 3 show that this feature is confirmed by the theoretical results, clearly and completely. Within the framework of the present theory, the mechanism for the oscillations can be analyzed as follows: When some of the wave numbers $p_\pm^z$ and $q_\pm^z$ change from real to imaginary, the transmission coefficient $T_{\uparrow\uparrow}$ will change from oscillating to damping with the barrier thickness. As a result, there will exist two kinds of integral regions for the channel of $T_{\uparrow\uparrow}$: On the first kind of region, $T_{\uparrow\uparrow}$ contains oscillating term $\cos[(p_+^z - p_-^z)d]$ or $\cos[(q_+^z - q_-^z)d]$. On the other kind of region, it contains neither $\cos[(p_+^z - p_-^z)d]$ nor $\cos[(q_+^z - q_-^z)d]$. For the other three channels, i.e., $T_{\uparrow\downarrow}$, $T_{\downarrow\uparrow}$ and $T_{\downarrow\downarrow}$, the situations are the same. We find through the numerical analysis that the oscillation of $R_P$ origins basically from the oscillating term of $\cos[(q_+^z - q_-^z)d]$ that belongs to the channel of $T_{\uparrow\uparrow}$. Similarly, the oscillation of $R_{AP}$ origins basically from the oscillating term of $\cos[(\tilde{p}_+^z - \tilde{p}_-^z)d]$ that belongs to the channel of $T_{\downarrow\uparrow}$. As mentioned above, the oscillating terms of $\cos[(q_+^z - q_-^z)d]$ and $\cos[(\tilde{p}_+^z - \tilde{p}_-^z)d]$ stand for the interference between the waves of $\mathbf{q}_+$ and $\mathbf{q}_-$, and that of $\tilde{\mathbf{p}}_+$ and $\tilde{\mathbf{p}}_-$, respectively. Physically, this interference arises from the diffraction of the tunneling electrons by the periodic potential of single-crystal barrier. That is to say, it is the coherent diffraction of electrons by the periodic potential that is actually the physical mechanism for the oscillations.

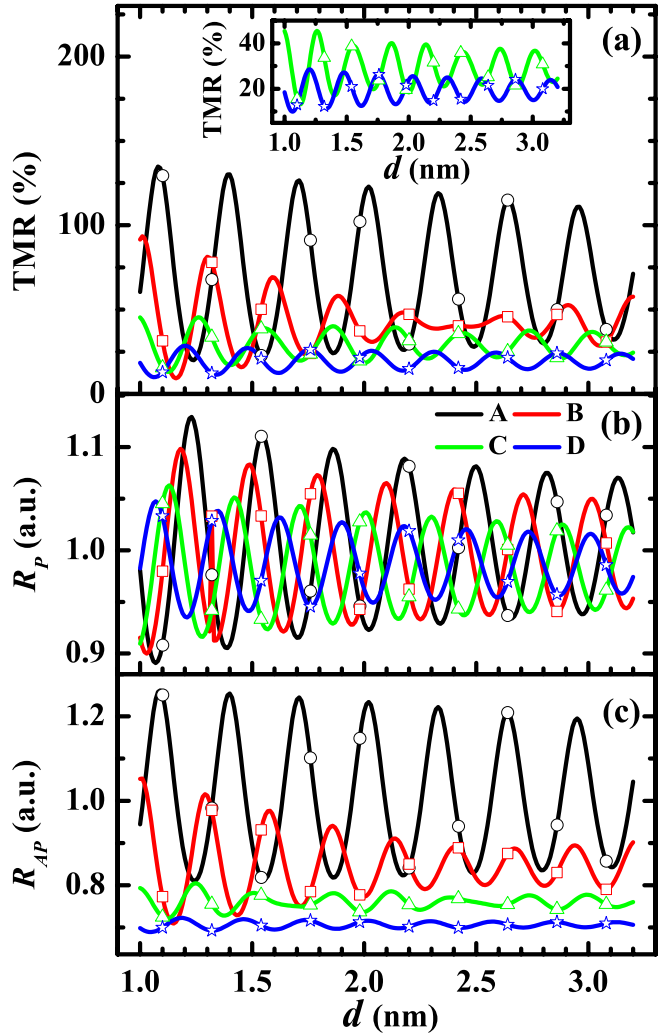

**Figure 2. Thickness dependence of TMR, $R_P$ and $R_{AP}$ with the variation of $\Delta$.** (**a**) TMR, (**b**) $R_P$ and (**c**) $R_{AP}$ as functions of barrier thickness $d$ where $K_h = 2.116 \times 10^{10} \, \text{m}^{-1}$, $v(\mathbf{K}_h) = 16 \, \text{eV}$ and $\mu = 11 \, \text{eV}$. The curves A, B, C, and D correspond to $\Delta = 10 \, \text{eV}$, $9 \, \text{eV}$, $8 \, \text{eV}$ and $7 \, \text{eV}$, respectively. The data corresponding to the discrete layers of the grating barrier are marked with open symbols: curve A is marked by open circle, curve B is marked by open square, curve C is marked by open triangle, and curve D is marked by open star. Here, the inset is an enlarged view of the curves C and D.

Now, we proceed to interpret the details of the oscillations. As mentioned above, the oscillations of $R_P$ and $R_{AP}$ origin from the terms of $\cos[(q_+^z - q_-^z)d]$ and $\cos[(\tilde{p}_+^z - \tilde{p}_-^z)d]$, respectively, and thus $q_+^z - q_-^z$ and $\tilde{p}_+^z - \tilde{p}_-^z$ are their frequencies of oscillation. With the changes in the incident wave vectors $\mathbf{k}_\uparrow$ and $\mathbf{k}_\downarrow$, both $q_+^z - q_-^z$ and $\tilde{p}_+^z - \tilde{p}_-^z$ will vary and thus form two frequency bands for $G_P$ and $G_{AP}$, respectively. Obviously, the periods of oscillations depend on the structures of the frequency bands, which can be described by the spectral densities $\mathcal{D}_P(\omega)$ and $\mathcal{D}_{AP}(\omega)$,

$$\begin{aligned}\mathcal{D}_P(\omega) &= \frac{e^2}{8\pi^3 \hbar} \int_0^{\pi/2} d\theta \int_{-\pi/4}^{\pi/4} d\varphi k_{F\uparrow} \sin(\theta)(q_+^z + q_-^z)\delta(\omega - (q_+^z - q_-^z)) \\ &\quad \times H((q_+^z)^2, (q_-^z)^2),\end{aligned} \tag{9}$$

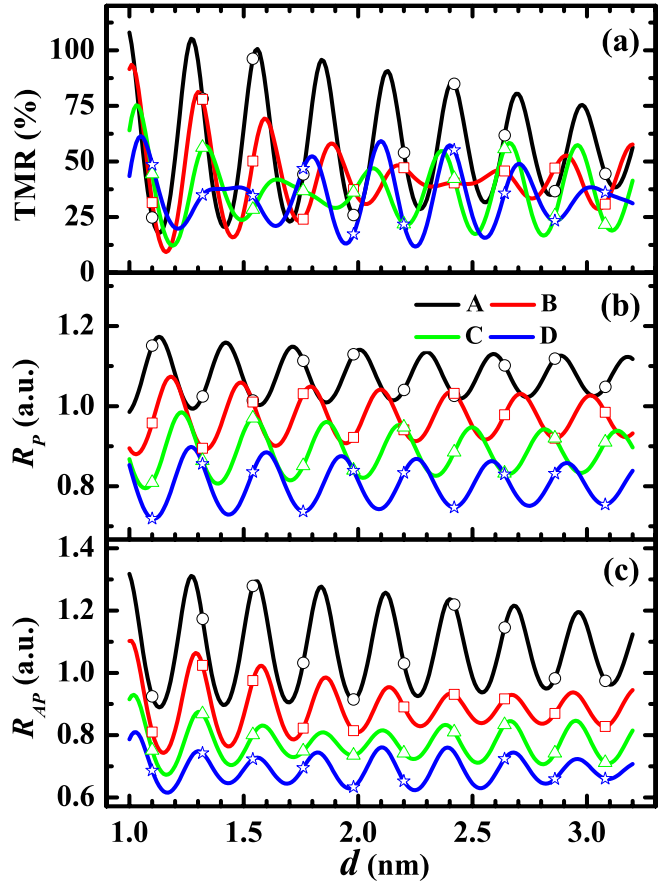

**Figure 3. Thickness dependence of TMR, $R_P$ and $R_{AP}$ with the variation of $\mu$.** (**a**) TMR, (**b**) $R_P$ and (**c**) $R_{AP}$ as functions of barrier thickness $d$ where $K_h = 2.116 \times 10^{10}\,\text{m}^{-1}$, $\nu(\mathbf{K}_h) = 16\,\text{eV}$ and $\Delta = 9\,\text{eV}$. The curves A, B, C, and D correspond to $\mu = 10\,\text{eV}$, $11\,\text{eV}$, $12\,\text{eV}$ and $13\,\text{eV}$, respectively. The data corresponding to the discrete layers of the grating barrier are marked with open symbols: curve A is marked by open circle, curve B is marked by open square, curve C is marked by open triangle, and curve D is marked by open star.

$$\mathcal{D}_{AP}(\omega) = \frac{e^2}{8\pi^3\hbar} \int_0^{\pi/2} d\theta \int_{-\pi/4}^{\pi/4} d\varphi\, k_{F\downarrow}\, \sin(\theta)(\tilde{p}_+^z + \tilde{p}_-^z)\delta(\omega - (\tilde{p}_+^z - \tilde{p}_-^z)) \\ \times H((\tilde{p}_+^z)^2, (\tilde{p}_-^z)^2), \qquad (10)$$

where $H(x)$ is the Heaviside function. The numerical results for $\mathcal{D}_P(\omega)$ and $\mathcal{D}_{AP}(\omega)$ are shown respectively in Fig. 4(a,b), which correspond to the cases of Fig. 2. The situations for Fig. 3 are similar.

Figure 4(a) demonstrates that $\mathcal{D}_P(\omega)$ can be approximated as a narrow peak plus a straight segment. As well-known, the peak will provide principally one frequency, say $\omega_0$, to $G_P$. In addition, it is easy to know by direct integration that the function with a finite high-frequency band and a linear spectral density will look as if it has merely two frequencies on the domain where its variable is quite large: one is the band bottom, the other is the band top. So does the present case, the straight segment will contribute two frequencies to $G_P$, but one of them is zero because $\mathcal{D}_P(\omega)$ vanishes at the right endpoint of the segment. The rest nonzero one, say $\omega_1$, is equal to the frequency of the left endpoint of the segment. In sum, there will exist mainly two frequencies for the oscillations of $G_P$, i.e., $\omega_0$ and $\omega_1$. However, as indicated by Fig. 4(a), $\omega_0$ and $\omega_1$ are very close to each other. This implies that the modulation frequency of the resultant oscillation is rather small. Or equivalently, the modulation period will be much longer than the basic period which is about $4\pi/(\omega_0 + \omega_1)$, as can be seen from Fig. 2(b). For the experiments up to now[16,18], the varying range of the barrier thickness is approximately 1.2 nm, which is too narrow for the modulation period to be observed. That explains why only one single period for $R_P$ has been reported by the experiments[16,18].

As to $\mathcal{D}_{AP}(\omega)$, it can be further simplified as follows: $\mathcal{D}_{AP}(\omega) = e^2\, \tilde{p}_+^z\, \tilde{p}_-^z (\tilde{p}_+^z + \tilde{p}_-^z)/[16\pi^2\hbar k_{F\downarrow} \cos(\theta)(\tilde{p}_+^z - \tilde{p}_-^z)]$ where $\tilde{p}_+^z$, $\tilde{p}_-^z$ and $\theta$ are all the functions of $\omega$. From this formula, one can easily know that, when $\Delta > \nu(\mathbf{K}_h)/2$, $\mathcal{D}_{AP}(\omega)$ will get a von Hove singularity, which is at the band top where $\theta = \pi/2$. The corresponding results are depicted in the curves A and B of Fig. 4(b). In this case, $\mathcal{D}_{AP}(\omega)$ can also be approximated as a peak plus a straight segment, the peak comes from the singularity. Of course, the singularity contributes one frequency, say $\omega_s$, to $G_{AP}$. The straight segment will, as before, contribute principally two frequencies, say $\omega_l$ and $\omega_r$, to $G_{AP}$ where $\omega_l$ and $\omega_r$ are equal to the frequencies of the left and right endpoints of the segment, respectively. Therefore, there exist now mainly three frequencies, $\omega_s$, $\omega_l$ and $\omega_r$. In the case of narrow band, e.g., the curve A, the three frequencies are nearly the same, so only one single period can be observed experimentally for $R_{AP}$. In the case of wide band, e.g., the curve B, only $\omega_s$ and $\omega_r$ are close to each other, but $\omega_l$ is far away, the three frequencies will be seen as two in the experiments. When $\Delta \leq v(\mathbf{K}_h)/2$, there is no von Hove singularity in $\mathcal{D}_{AP}(\omega)$. In this case, $\mathcal{D}_{AP}(\omega)$ is nearly a linear function of $\omega$, as shown in the curves C and D of Fig. 4(b). If $\Delta = v(\mathbf{K}_h)/2$, $\mathcal{D}_{AP}(\omega_M) = e^2 \tilde{p}_+^z /(16\pi^2\hbar) > 0$ where $\omega_M$ is the maximal frequency of the band. Otherwise, $\mathcal{D}_{AP}(\omega_M) = 0$. This implies that there are mainly two frequencies in $G_{AP}$ when $\Delta = v(\mathbf{K}_h)/2$, and one frequency when $\Delta < v(\mathbf{K}_h)/2$. All in all, there can exist one or two periods for $R_{AP}$, which is in agreement with the experiments[16,18].

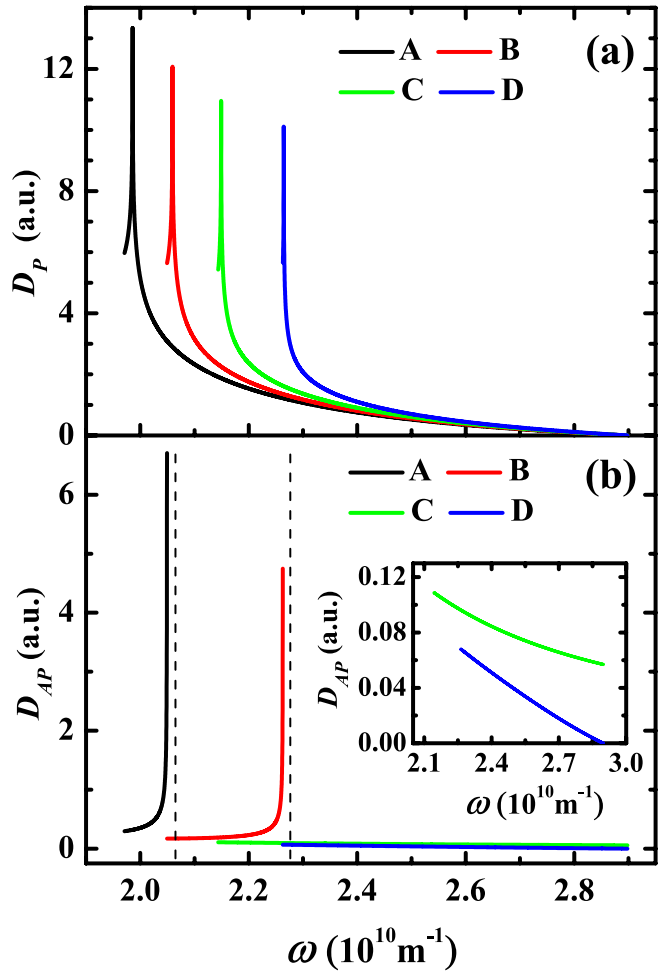

**Figure 4. Spectral densities.** (**a**) $\mathcal{D}_P$ and (**b**) $\mathcal{D}_{AP}$ as functions of the oscillating frequency $\omega$ where $K_h = 2.116 \times 10^{10}$ m$^{-1}$, $v(\mathbf{K}_h) = 16$ eV and $\mu = 11$ eV. The curves A, B, C, and D correspond to $\Delta = 10$ eV, 9 eV, 8 eV and 7 eV, respectively. Here, the dashed lines denote the von Hove singularities, the inset is an enlarged view of the curves C and D.

Finally, the TMR, as the ratio of $R_P$ and $R_{AP}$, will exhibit mainly one or two periods. Within the range of present parameters, the basic period of TMR distributes on the interval of [0.28 nm, 0.31 nm], which is in agreement with the experimental data, from 0.28 nm to 0.32 nm[10,16–19]. In particular, the numerical results show that the basic periods of $R_P$, $R_{AP}$ and TMR are nearly the same, with the relative error less than 4.1%, which is also in agreement with the points of view of refs 16,18.

As for the amplitudes, equation (9) indicates that the amplitude of $G_P$ will increase with the sum $q_+^z + q_-^z$. From equation (5), it can be seen that the sum $q_+^z + q_-^z$ increases with the parameter $\mu + \Delta$. Therefore, the amplitude of $G_P$ will increase with the parameter $\mu + \Delta$. Numerical analysis shows that, the larger the amplitude of $G_P$, the larger the amplitude of $R_P$, which means that, the larger the parameter $\mu + \Delta$ is, the larger the amplitude of $R_P$ will be. Similarly, the smaller the difference $\mu - \Delta$, the larger the amplitude of $R_{AP}$. Those properties are shown clearly in Figs 2 and 3. Experimentally, the amplitude of TMR is about 100% in the case with only one period, and 20% ~ 40% in the case with another long period[16,18,19]. Evidently, those facts are included in the present theoretical results, as displayed in Figs 2 and 3.

Secondly, we would like to investigate the influence of the parameter $v(\mathbf{K}_h)$. The theoretical results are shown in Fig. 5 where $v(\mathbf{K}_h)$ is set sequentially as 12 eV, 16 eV and 20 eV while $\mu$ and $\Delta$ are fixed to be 11 eV and 10 eV, respectively. Figure 5(a) demonstrates that both the amplitude and period of $G_P$ increase with decreasing $v(\mathbf{K}_h)$. That is because the sum $q_+^z + q_-^z$ in equation (9) will increase, but the difference $q_+^z - q_-^z$ in the interference term of $\cos[(q_+^z - q_-^z)d]$ decrease, with decreasing $\nu(K_h)$. Similar statements and analyses are valid for $G_{AP}$, too. The distinction between $G_P$ and $G_{AP}$ lies in their direct-current components: For $G_P$, the direct-current component is weakly dependent on the variation of $\nu(K_h)$. Nevertheless, that of $G_{AP}$ increases strongly with the increase of $\nu(K_h)$. As a consequence of those results, the amplitude and period of TMR will both increase with decreasing $\nu(K_h)$, which is depicted in Fig. 5(c). As has been pointed out above, the parameter $\nu(K_h)$ is approximately proportional to the energy gap of bands. Therefore, if the MgO barrier is replaced by a semiconductor with smaller (or larger) energy gap, both the amplitude and period of TMR will become larger (or smaller). This demonstrates that the oscillation of TMR can be modulated through the gap of barrier, both in frequency and in amplitude. Here, it should be emphasized, in particular, that the barrier with a small gap is especially favourable for the TMR to get a large amplitude. Now, consider the layered semiconductor $MoS_2$, its intralayer gap is about 1.9 eV[38]. So, if a MTJ has $MoS_2$ as the barrier, its TMR is very believable to acquire a much higher amplitude than those of the MTJs with MgO barrier, which can be used to verify the present theory in the future.

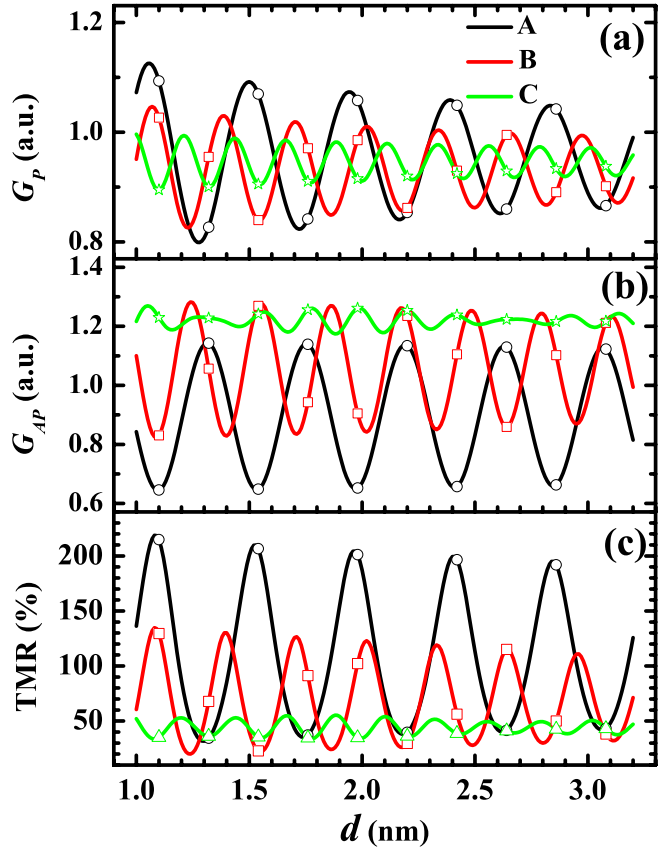

**Figure 5. Thickness dependence of $G_P$, $G_{AP}$ and TMR with the variation of $\nu(K_h)$.** (**a**) $G_P$, (b) $G_{AP}$ and (**c**) TMR as functions of barrier thickness $d$ where $K_h = 2.116 \times 10^{10}\,m^{-1}$, $\mu = 11\,eV$ and $\Delta = 10\,eV$. The curves A, B, and C correspond to $\nu(K_h) = 12\,eV$, $16\,eV$ and $20\,eV$, respectively. The data corresponding to the discrete layers of the grating barrier are marked with open symbols: curve A is marked by open circle, curve B is marked by open square, and curve C is marked by open triangle.

Thirdly, we shall discuss the TMR for the MTJ using half-metallic electrodes where the spin is fully polarized. In conventional theories[1,8,26–30], the TMR will become nearly infinite near zero bias and at low temperature when both electrodes are fully spin-polarized, just as pointed out in ref. 39. Physically, that is because, in conventional theories, the energy of each tunneling electron is conserved and thus there will exist no spin-down state for the incident spin-up electron to occupy, i.e. $G_{AP} = G_{\uparrow\downarrow} = 0$. Unfortunately, this result is contrary to the experiments up to now[17,40–42] where the TMR is much far away from infinity. In two-beam approximation, the electron waves pass directly from the barrier into the lower electrode, unchanged except that the barrier wave vectors become the lower-electrode wave vectors[31]. Evidently, such treatment belongs physically to the so-called sudden approximation[43]. As a result, the energy of the tunneling electron can be non-conserved. For example, the transmitted component waves with $\mathbf{p}_+$ and $\mathbf{q}_+$ acquire an energy of $\nu(K_h)$, and those with $\mathbf{p}_-$ and $\mathbf{q}_-$ lose an energy of $\nu(K_h)$, as can be seen from the equations (3–5). Due to this nonconservation, the incident spin-up electrons can transit into the spin-down band of the lower electrode, with a probability of 50% or so, if $\nu(K_h) > \Delta - \mu$, which is depicted in Fig. 6. That implies $G_{\uparrow\downarrow} \neq 0$ if $\nu(K_h) > \Delta - \mu$. In other words, when $\nu(K_h) > \Delta - \mu$, the TMR will still be finite other than infinite, i.e., $TMR = G_{\uparrow\uparrow}/G_{\uparrow\downarrow} - 1 < +\infty$, even if both the upper and lower electrodes are half-metallic. On the contrary, if $\nu(K_h) < \Delta - \mu$, the energy will be insufficient for all the out-going component waves, the incident spin-up electrons can not transit into the spin-down band of the lower electrode yet, so $G_{\uparrow\downarrow} = 0$ and $TMR = +\infty$. In a word, within the framework of the present theory, the TMR can be finite or infinite for the MTJ using both half-metallic electrodes, which depends on whether $v(\mathbf{K}_h) > \Delta - \mu$ or not. As stated above, the energy gap of MgO is 8 eV, it is rather large, and thus can satisfy quite easily the condition: $v(\mathbf{K}_h) > \Delta - \mu$. That may be the physical reason why the TMR for the experiments up to now[17,40–42] is much far away from infinity. Of course, if the barrier could be fabricated by the single crystal with an energy gap small enough, the TMR would tend to infinity. As such candidates, people can naturally consider the common semiconductors, such as Ge, InN, InAs, $MoS_2$, etc. Here, we would, in particular, suggest the $\alpha$ graphite[44,45] which are stacked by graphene planar layers. As well-known, the graphene layers are linked just through a weak van der Walls interaction, so the interlayer conductivity is much little, and thus can be neglected and regarded as insulating. Observe that graphene is a zero-gap semiconductor, with the conduction and valence bands meeting at the Dirac points. If a thin film of the multilayer of graphene cleaved out of $\alpha$ graphite is used as the barrier, one has $v(\mathbf{K}_h) = 0$ immediately. Therefore, the condition $v(\mathbf{K}_h) < \Delta - \mu$ can always be satisfied, which means that, if a MTJ forms of two half-metallic electrodes and a graphite barrier, its TMR can approach infinity. It is both a prediction to verify the present theory and a suggestion for manufactories to assemble MTJs in a simple way. This simple way is also suitable for the $MoS_2$ nano-film, which has been discussed above.

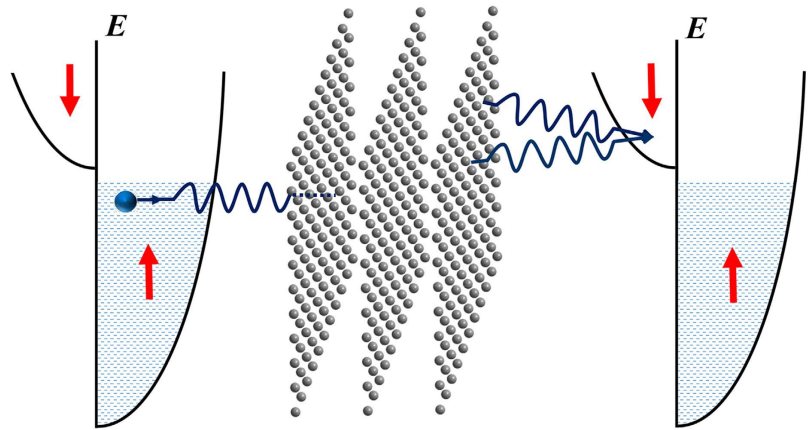

**Figure 6. The diagrammatic sketch of the diffraction process for the antiparallel case where the two electrodes are both half-metallic.** Here, $v(\mathbf{K}_h) > \Delta - \mu$, the incident spin-up electron can transit into the spin-down band of the lower electrode.

Finally, there is another radical difference between the usual theories[1,8,26–30] and the present one. For the former, the transmission coefficient decreases exponentially with the barrier width so that the tunneling will disappear thoroughly as the barrier gets quite thick, e.g., thicker than 50 nm. For the latter, the tunneling electrons will move within a periodic potential constructed by the barrier. As well-known, the waves will be formed into band structure within a periodic potential, and thus can be transported very far away. Back to the present case, according to the two-beam approximation, the tunneling electrons will form two narrow bands within the periodic barrier, which arise respectively from the states near the two Fermi surfaces of the upper FM electrode. Just as in the Laue case of electron diffraction, it is not difficult to know that there always exist a large amount of travelling waves in those two bands even if the thickness of the periodic barrier goes into the range of 1 $\mu$m ~ 1 mm. This means that the oscillating $G_P$, $G_{AP}$, and TMR can be maintained for the MTJ with a barrier thicker than 1 $\mu$m. Nowadays, single-crystal semiconductor sheets of thickness ranging from 1 $\mu$m to 1 mm can be readily made in industries. Those analyses suggest that the width of crystalline barrier can get rid of the confinement of nanoscale and goes into microscale, not theoretically but also technologically. As a consequence, a MTJ with the periodic grating barrier of thickness within 1 $\mu$m ~ 1 mm can be simply synthesized by direct sandwiching of a single-crystal semiconductor sheet between two FM electrodes rather than by MBE growing of the barrier. It is a novel design of MTJ that will work on electron diffraction, take advantage of the coherence of tunneling electron waves, be more appropriate for the materials with a layered, planar structure, such as $MoS_2$[38] and graphite[44,45], and, in particular, be most feasible for industries.

In summary, this paper has developed a new spintronic theory. Physically, it is founded on the optical diffraction theory, and thus particularly suitable for the MTJs with a single-crystal barrier which plays a role of periodic optical grating. In consequence, there will appear strong coherence among the diffracted waves of tunneling electrons. It is just this coherence that is responsible for the oscillations of the $R_P$, $R_{AP}$ and TMR with the barrier thickness. As such, the theory can well interpret the experiments on the MgO-based MTJs: (1) There exists nearly a common period among the $R_P$, $R_{AP}$ and TMR in all the cases. (2) The $R_{AP}$ and TMR can show another long period of oscillation in some cases. (3) The amplitude of TMR is about 100% in the case that there is only one period, and is 20% ~ 40% in the case that there is another long period. In particular, the theory shows that, the smaller the in-plane energy gap of the barrier material, the larger the amplitude and period of the TMR will be, which implies that $MoS_2$ and graphite would be more favorable than MgO as a barrier for MTJs. Besides, the present theory can explain the puzzle why the TMR is still far away from infinity when the two electrodes are both half-metallic, and suggests further a possible way to enhance the TMR so that it could tend to infinity. Finally, a future-generation MTJ is proposed, it is a triplet of two FM electrodes plus a single-crystal barrier, the three members can be fabricated separately, and then assembled together and encapsuled in a unit, the barrier can vary from nanoscale to microscale in width. We believe that such an architecture can take advantage of the coherence of tunneling electrons, is especially appropriate for the layered materials, e.g., $MoS_2$ and graphite, and most feasible for industries.

By the way, there are many important and interesting effects that are highly correlated with the oscillations of TMR, such as bias effect[10,18], temperature effect[9], impurity effect[46], etc. The studies of these effects are in progress and will be published in future.

### Acknowledgements
This work is supported by the State Key Program for Basic Research of China (2014CB921101), the Nature Science of Foundation of Jiangsu province (BK20130866), the National Natural Science Foundation of China (61574079, 51471085, 51331004), the University Nature Science Research Project of Jiangsu province (14KJB510020), the NUPTSF (NY213025, NY215083).

### Author Contributions
M.X. and H.F. conceived the project. H.F. carried out the calculations. M.X. led the analysis. H.F., M.X. and J.D. wrote the paper. H.F., W.R. and Z.T. contributed to the preparation of the figures. M.X. supervised the study.

### Additional Information
**Competing financial interests:** The authors declare no competing financial interests.

**How to cite this article**: Fang, H. *et al.* Magnetic coherent tunnel junctions with periodic grating barrier. *Sci. Rep.* **6,** 24300; doi: 10.1038/srep24300 (2016).